# The new model of fitting the spectral energy distributions of Mkn 421 and Mkn 501

GAO XiaoYan[1†]   WANG JianCheng[1]   YANG JianPing[1]

[1] National Astronomical Observatories /Yunnan Observatory, Chinese Academy of Sciences, Kunming 650011, China

**The spectral energy distribution (SED) of TeV blazars has a double-humped shape that is usually interpreted as Synchrotron Self Compton (SSC) model. The one zone SSC model is used broadly but cannot fit the high energy tail of SED very well. It need bulk Lorentz factor which is conflict with the observation. Furthermore one zone SSC model can not explain the entire spectrum. In the paper, we propose a new model that the high energy emission is produced by the accelerated protons in the blob with a small size and high magnetic field, the low energy radiation comes from the electrons in the expanded blob. Because the high and low energy photons are not produced at the same time, the requirement of large Doppler factor from pair production is relaxed. We present the fitting results of the SEDs for Mkn 501 during April 1997 and Mkn 421 during March 2001 respectively.**

## 1 Introduction

Blazar is a subclass of active galactic nuclei (AGNs) that oriented at a small angle with respect to the line of sight. They always emit non-thermal radiation from radio to gamma-rays, even up to TeV energy[1]. The spectral energy distribution (SED) of blazars have two peaks: the first one is in the radio / UV range even up to soft X-ray range and the second one is in the X-ray /gamma-ray range.

The first peak component of the SED is usually attributed to synchrotron emission from relativistic electrons. The leptonic and hadronic models are two main scenarios to explain the second peak component. In the leptonic model, gamma-rays are produced by electron inverse-Compton (IC) scattering of photons which come from internal or external emission region. Hadronic model assumes that the ultra-high energy protons lead to gamma ray emission after interaction and decay of secondary particles[2-4]. Two subclass of hadronic model are Proton-Initiated Cascade (PIC) model and Synchrotron Proton Blazar (SPB) model. PIC model is pion photoproduction by energetic protons with subsequent synchrotron-pair cascades initiated by decay products (photons and $e^{\pm}$) of the mesons[5-6]. The SPB model is another attractive possibility for production of high-energy γ-ray, where high -energy protons are accelerated and emit subsequent synchrotron radiation. The SPB model requires ultra high energy protons and strong magnetic fields in the small emission region[7].

The discovery of strong TeV variability in two blazers, PKS 2155-304[8] and Mrk 501[9], on timescales as short as a few minutes, implies a compact emission region that moves with a large bulk Lorentz factor of $\Gamma > 50$ towards the observer assuming a homogeneous one-zone model. However, such high values of the bulk Lorentz factor are in contradiction with constrains derived from other observational evidence.[10,11] Furthermore, one-zone models are unable to fit the entire spectrum, the low-energy synchrotron component being long variability timescale which is generally attributed to large emitting region.

In the paper, we present a new model, unifying small and large scales emission region: we consider that the accelerated protons in the original blob with a small size dominate the high energy TeV emission and the electrons in the expanded blob contribute the low energy emission and the tail of high energy radiation. The present model only concentrates on the physical parameters of the relativistic blob and is described in Section 2. We show its application to the cases of Mrk 421 and Mrk 501 in Section 3 focusing on different states.

## 2. The Model

We assume the high energy emission region to be a blob near the black hole with relativistic protons, electrons and a uniform field. We need three parameters to describe the blob. They are the radius $R_i \leq \delta c \Delta t_i$, where $\Delta t_i$ is the timescale of high energy radiation variability, the uniform magnetic field $B_i$ and the Doppler factor $\delta$. We assume that the electrons and protons in the blob are accelerated to form the standard energy distribution follow as:

$$N(\gamma) = N_0 \gamma^{-\alpha} \exp(-\gamma/\gamma_{\max}), \qquad (1)$$

where $N_0 \in (N_{p0}, N_{e0})$ presents the number density of protons or electrons, $\alpha \in (\alpha_p, \alpha_e)$ and $\gamma_{\max} \in (\gamma_{p,\max}, \gamma_{e,\max})$ are their spectral index and the cutoff energy. In order to produce TeV emission dominated by the proton synchrotron, the magnetic field is limited by[7]

$$B_i \geq 80 \left(\frac{R}{10^{15} cm}\right)^{-2/3} \eta^{1/3} G, \qquad (2)$$

The cutoff energy $\gamma_{p,\ max}$ and $\gamma_{e,\ max}$ are given by[7]

$$\gamma_{p,\max} \leq 2.1 \times 10^9 \left(\frac{R}{10^{15} cm}\right)^{1/3} \eta^{-2/3}, \qquad (3)$$

$$\gamma_{e,\max} \leq 1.1 \times 10^6 \left(\frac{R}{10^{15} cm}\right)^{1/3} \eta^{-2/3}, \qquad (4)$$

and the observed cutoff energies of the proton and electron synchrotron spectra are determined by

$$\nu_p = 0.3 \eta^{-1} \delta \text{ TeV}, \qquad (5)$$

$$\nu_e = 0.16 \eta^{-1} \delta \text{ GeV} \qquad (6)$$

where $\eta$ is the so-called gyro-factor charactering the rate of proton acceleration and remains a rather uncertain model parameter. In the case of diffusive shock acceleration in the blazar jets the parameter $\eta$ is expected to be larger than $10$[12].

The coefficient of proton synchrotron emission is

$$j_s'(\nu_p') = \frac{1}{4\pi} \int_{\min}^{\max} N_p(\gamma) P(\nu_p', \gamma) d\gamma, \qquad (7)$$

where the primed parameters are in blob frame, $\nu_p' = \delta^{-1} \nu_p$, $P(\nu_p', \gamma)$ is the spectral distribution of synchrotron radiation emitted by a proton of energy $\gamma$ and is given by

$$P(\nu_p', \gamma) = \frac{\sqrt{2} e^3 B}{m_p c^2} F(x). \qquad (8)$$

where $x = \nu_p'/\nu_c$ and $F(x) = x \int_x^\infty dx K_{5/3}(x)$; $K_{5/3}(x)$ is the modified Bessel function of 5/3 order.

$$v_c = \sqrt{\frac{3}{2}} \frac{eB\gamma^2}{2\pi m_p c}$$ which is for the magnetic field distributed isotropically.

The emission intensity in a uniform field is given by[11-12],

$$I'_s(v'_p) = \frac{j'_s(v'_p)}{k'(v'_p)}(1 - \frac{2}{\tau_{rr}^2}[1 - e^{-\tau_{rr}}(\tau_{rr}+1)]), \quad (9)$$

where $K(v'_p)$ is the absorption coefficient of gamma-ray photons by pair production, and $\tau_{\gamma\gamma} = 2k'(v'_p)R$ is the optical depth in the blob. There are two possible soft photons for pair production: the radiation intrinsic to the blob and the external ambient radiation. We assume that $N_{e0} \ll N_{p0}$ and the absorption of the photon produced by the electrons to gamma-rays can be ignored to keep $\tau_{\gamma\gamma} \leq 1$.

The observed SED of the proton synchrotron is given by

$$v_p F_s(v_p) = \pi \frac{R^2}{d_l^2} \delta^3 (1+z) v_p I'_s(v'_p). \quad (10)$$

We assume that the blob moves away the black hole with a constant $\delta$ and expands to the radius $R_f$. The magnetic field in blob decreases as

$$B_f = B_i \left(\frac{R_i}{R_f}\right)^2, \quad (11)$$

the proton synchrotron is restrained and the high energy electrons produced by acceleration mechanisms or proton-photon cascades dominates the radiation of the blob. The electrons emit the synchrotron radiation to contribute the low-energy spectrum of the SED and inverse Compton (SSC) radiation to supply the high-energy spectrum. We also adopt the standard electron energy distribution with the form:

$$N_f(\gamma) = N_{f0} \gamma^{-\alpha_f} \exp(-\gamma/\gamma_{f,max}), \quad (12)$$

and take the variability timescale $\Delta t_f$ of low energy radiation to limit the radius $R_f \leq \delta c \Delta t_f$. We then calculate the emissive spectra of the electrons to fit the observed SED of low energy component.

Calculating the gamma-ray spectrum we have considered the absorption of Intergalactic Infrared Background (IIB) for gamma-ray emission. We adopt the absorption coefficient of the infrared intergalactic radiation with a new empirical calculations derived by Stecker & De Jager (1998)[13].

## 3. Fitting the SED of Mrk421 and Mrk501

We get the simultaneous observation data of Mrk 421 observed in Match 19 and Match 22-23 2001 and Mrk501 observed in Match 7 and 16 1997. Giebels et al.[14] have modified the SED of Mrk421 in 2001 Match using IC model, while Pian et al.[15] fitted the SED of Mrk501 with inhomogeneous SSC model. But they cannot fit the high energy spectra very well. There always exists a turn up tail. We use new model to fit their SEDs. The finally fitted spectra for two epochs of Mrk421 and Mrk 501 are shown in figure 1 to figure 4.

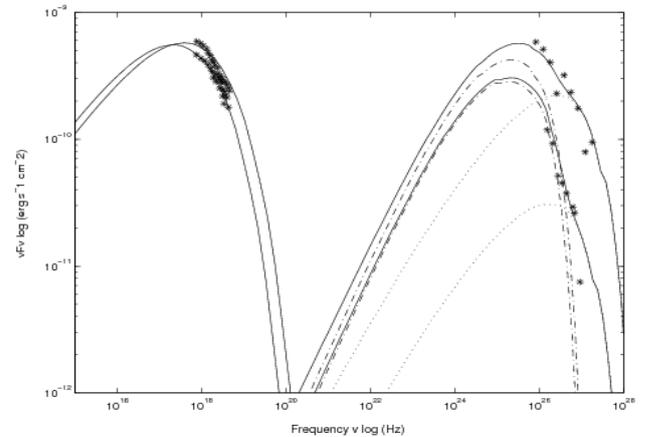

Figure 1: Fitted the Mrk421 with summation of SPB model (dash-dotted line) and SSC (dotted line) for supplement for the epoch of Match 19 (high state) and Match 22-23 (low state) in 2001. The simultaneous VHE (CAT), X-ray (RXTE)[11]. The solid line in lower energy band is due to synchrotron of electrons.

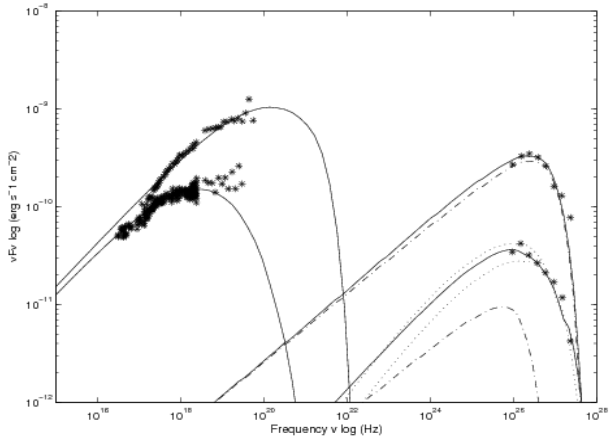

Figure 2: Fitted the Mrk501 with SPB model (dash-dotted line) and SSC (dotted line) for supplement for the epoch of April 7 (low state) and April 16 (high state)1997. The solid line in lower energy band is due to synchrotron of electrons. The simultaneous data is taken by the Beppo-SAX instrument [12]

The values of magnetic field B, emitting radius R and Doppler factor δ are shown in table 1.

Table 1

| | | March 19 for Mrk 421 | March 22-23 for Mrk421 | April 7 for Mrk501 | April 16 for Mrk501 |
|---|---|---|---|---|---|
| R | $R_f$ | $2.1 \times 10^{16}$ | $1.3 \times 10^{16}$ | $4.7 \times 10^{17}$ | $7.3 \times 10^{17}$ |
| | $R_i$ | $6.2 \times 10^{14}$ | $4.1 \times 10^{14}$ | $3.4 \times 10^{16}$ | $2.3 \times 10^{16}$ |
| B | $B_f$ | 0.2 | 0.3 | 0.4 | 0.1 |
| | $B_i$ | 237 | 312 | 76 | 99 |
| δ | | 23 | 15 | 13 | 9 |
| $L_h/L_l$ | | 0.99 | 0.26 | 0.32 | 0.38 |
| $N_{p0}/N_{e0}$ | | $9.9 \times 10^4$ | $2.8 \times 10^4$ | $7.9 \times 10^4$ | $4.0 \times 10^4$ |

We can also estimate the ratio of proton and electron numbers from the two peak luminosities of the observed SED. The peak luminosities are given by $L_h \propto N_{p0} B_i^2 R_i^3 m_p^{-2}$ for the protons and $L_l \propto N_{e0} B_f^2 R_f^3 m_e^{-2}$ for the electrons respectively. We can get,

$$\frac{N_{p0}}{N_{e0}} = \left(\frac{L_h}{L_l}\right)\left(\frac{B_i}{B_f}\right)^{-2}\left(\frac{R_i}{R_f}\right)^{-3}\left(\frac{m_p}{m_e}\right)^2 = \left(\frac{L_h}{L_l}\right)\left(\frac{R_i}{R_f}\right)\left(\frac{m_p}{m_e}\right)^2$$

(13)

For the two epochs of Mrk421 and Mrk501, the time scale $\Delta t_f$=15 minutes for high energy emission and $\Delta t_f$=1 day respectively for March 19 and 21/22 of Mrk421 and May7 and 16 of Mrk501. The values $\frac{L_h}{L_l}$ are from observation. We can get the ratio, $\frac{N_{p0}}{N_{e0}}$ which are also show in table 1.

### 4. Conclusions and Discussion

We have fitted the high energy SED of the simultaneous data of Mrk 421 and Mrk 501 with the SPB model and the electron SSC model. Because the high energy emission is firstly produced by the protons in the small size of the blob and escapes the emission region, the gamma-rays will not be absorbed by the soft photons of the SED's low energy component which is produced by the electrons in the expanded blob. The high and low energy photons are not produced in the same time. The large Doppler factor from the limit of pair absorption can be relaxed.

We take into account the SSC contribution of the electrons to the gamma-rays in the expanded blob. From the fitting results we think that the tail of high energy band could be the SSC contribution of the electrons. We also estimate the ratio between protons and electrons.

The low state SED in X-ray band shown in Fig 2 is not fitted very well. In fact, these data points have big error bars due to fast flux variation. They contribute small right-weight in fitting the SED.

Our model does raise some questions in producing the observed nearly simultaneous X-ray/γ-ray variations of Mrk 421[16,17]. However, there have been regular multiwavelength campaigns for Mrk421 in the last several years. These campaigns reveal a rather loose correlation between the X-ray and TeV γ-ray fluxes[18,19]. For Mrk501, the multiwavelength observations lack a sufficiently long baseline to make a quantitative assertion about the statistical significance of a X-ray/TeV correlation[20]. There has even been evidence of an "orphan" TeV flare for the blazar 1ES 1959+650[21], a transient γ-ray event that was not accompanied by an obvious X-ray flare in simultaneous data. The key idea of the model is to decompose the TeV Blazar spectrum in different variability timescales and outburst states. The high-energy part of the spectrum, coming from small-scale regions, is considered to be one of quiescent and flaring states. On the other hand, the low-energy part is a convolution over a large scale regions including the past history of the blob: it is thus a spectral state mixing quiescent and flaring states, and shows a complex X-ray/TeV correlation. These possibilities lend added motivation to future observations and especially to simultaneous multiband observations.

*We acknowledge the financial supports from the National Natural Science Foundation of China 10673028 and 10778702, and the National Basic Research Program of China (973 Program 2009CB824800).*